\documentclass[11pt,twoside]{article}
\usepackage{newpasp,epsf}
\def\edcomment#1{\iffalse\marginpar{\raggedright\sl#1\/}\else\relax\fi}
\reversemarginpar

\begin{document}
\title{The nature of jets: evidence from circular polarization observations}
\author{John F. C. Wardle and Daniel C. Homan}
\affil{Physics Department, Brandeis University, Waltham, MA 02454, U.S.A.}
%\author{Ima Co-Author}
%\affil{The Name of My Institution, The Full Address of My Institution}

\begin{abstract}
We review recent observations of circularly polarized radiation from AGN made
with the VLBA and with the ATCA. We also discuss briefly the detections of 
the Galactic sources Sag A* and SS433. The origin of the circular 
polarization is still an open question in most cases, and we discuss four 
possible mechanisms.
Detectable circular polarization is a common property of quasars, but not of
radio galaxies, and is always associated with the compact core. There is
growing evidence that the {\em sign} of the circular polarization stays the 
same over at least two or three decades in time, suggesting it is a fundamental
property of the jet.

\end{abstract}

\section{Introduction}

Circular polarization (CP) measurements of extragalactic radio sources have
long enjoyed the dubious reputation of being challenging to carry out and 
hard to interpret. After a period of activity in the 1970's and early 80's
(summarized by Weiler \& de Pater 1983) there have been few new observations
until recently.

The renewed interest stems from two developments. First, the Australia
Telescope Compact array (ATCA) can now measure CP with a limiting accuracy
as low as $0.01\%$ (Rayner, Norris \& Sault 2000). This remarkable precision
is because the array was specifically designed to measure CP, and uses 
orthogonal {\em linearly} polarized feeds at each antenna. Second, the 
excellent performance of the VLBA has permitted the detection of CP in more 
than 20 AGN (Wardle et al. 1998, Homan \& Wardle 1999, Homan, Attridge \&
Wardle in preparation). Because the VLBA is equipped with {\em circularly} polarized
feeds, the limiting accuracy is $\sim 0.1\%$, some ten times worse than ATCA.
In compensation, the VLBA has about one thousand times higher resolution. 
Most measurements have been of AGN, but we
note that CP has also been detected in the Galactic sources Sag A$^*$
using the VLA (Bower, Falke \& Backer 1999) and SS433 using ATCA (Fender et
al. 2000).

The promise of CP measurements is that they may permit, depending on the
emission mechanism, direct measurement of the magnetic field strength and
its polarity, the true magnetic flux carried by a jet (and hence generated at
the central engine), the low energy cut-off in the relativistic electron 
spectrum, and the composition of the radiating plasma - whether
it is predominantly an electron-proton plasma or an electron-positron plasma,
or a mixture of the two.
None of these can be determined directly from measurements of the first
three Stokes parameters (total intensity, I, and linear polarization, Q and U).
The fourth Stokes parameter, V, corresponding to the circularly polarized
component of the radiation field, therefore offers a new way of measuring
fundamental physical properties of AGN, and also the smaller scale jets in
Galactic X-ray binaries.

\section{Circular polarization observations of AGN}

Following the VLBA detection of CP in the quasar 3C\,279 at 15 GHz (in which 
we argued that the jets were predominantly an electron-positron plasma,
Wardle et al. 1998), we published extensive 15 GHz observations of CP in four
AGN (3C\,84, PKS 0538+134, 3C\,273, 3C\,279; Homan \& Wardle 1999). This was 
part of a monitoring program in which 
sources were observed every two months for a year. In each source that was
detected (4 out of 13 $\simeq 32\%$, at fractional polarization levels, 
$m_C$, from $0.3$ to $1.0\%$), the CP signal was somewhat variable but
present throughout the year (and with the same sign), setting a minimum
time scale for large changes of CP in AGN. The CP signal was always
associated with the VLBI core, but often not coincident with it (e.g. 3C\,279).
Another clear example is 3C\,273 (see Figure 1), where the CP signal first 
appeared in the 
core concurrent with the appearance of significant linear polarization. It 
then moved away from the core and down the jet, coincident with a new moving
component in total intensity (Homan \& Wardle 1999). This behavior shows that
the CP is intrinsic to the source (as opposed to a foreground effect or a 
calibration error). It is most prominent where the optical 
depth is of order unity (as expected on theoretical grounds, Jones \& O'Dell 
1977, Jones 1988), and at least in these two quasars is associated with new 
components or shocks emerging from the optically thick base of the jet.

\begin{figure}
\plotfiddle{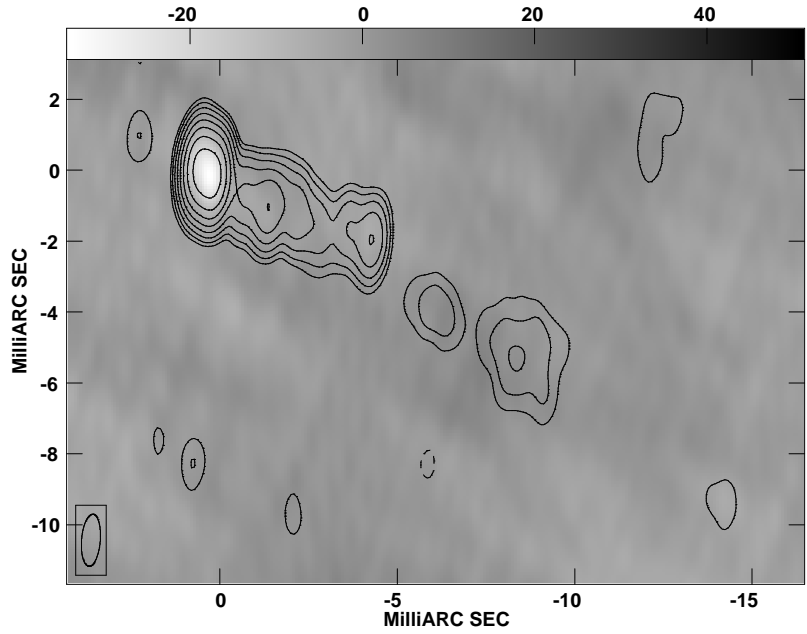}{3in}{0}{57}{57}{-190}{-120}
%\plotfiddle{j1130.eps}{1in}{-90}{25}{25}{0}{190}
%\begin{center}
%$\epsscale{0.7}
%\plotone{3c273.eps}
%\epsfig{file=3c273.eps,width=4in}
%\epsfig{file=j1130.eps,angle=-90,width=4in}
%\end{center}
%\vspace{-0.7in}
\caption{3C\,273 at 15 GHz, epoch 1996.74, showing circular polarization 
(grey scale) and total intensity (factors of 2 contours). The peak CP flux is 
$V = -37$ mJy, $V/I = -0.5\%$, located on a
new component ($I = 7.8$ Jy) emerging from the core region. }
\end{figure}

\begin{figure}
%\plotfiddle{3c273.eps}{1in}{0}{32}{32}{-200}{-107}
\plotfiddle{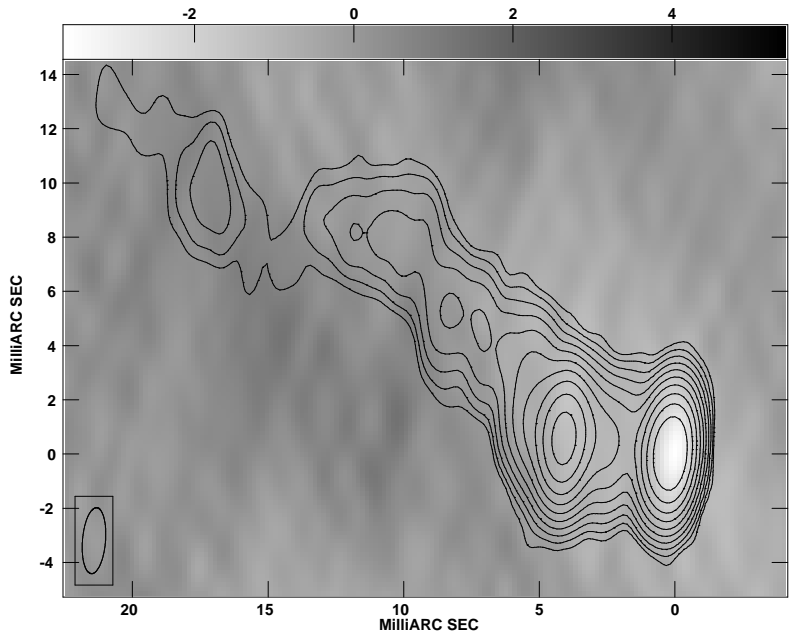}{3in}{-90}{45}{45}{-190}{250}
%\begin{center}
%\epsfig{file=3c273.eps,width=4in}\\
%\epsfig{file=j1130.eps,angle=-90,width=4in}
%\plotone{j1130.eps}
%\end{center}
\caption{PKS 1127-114 at 8 GHz, epoch 1998.01, showing circular polarization 
(grey scale) and total intensity (factors of 2 contours). The peak CP flux is 
$V = -3.2$ mJy, $V/I = -0.3\%$, coincident with the core ($I = 1.2$ Jy).}
\end{figure}

A VLBA survey of 40 blazars at 5 GHz (Attridge et al. 1998) has been 
reanalyzed for
circular polarization (Homan, Attridge \& Wardle, in preparation). We 
detected CP in 13 sources ($\simeq 33\%$), with
$m_C$ between $0.14$ and $0.46\%$, somewhat lower than those measured 
at 15 GHz. This sample is large enough to look for obvious correlations.
We detected CP in 11 out of 29 quasars, but only 2 out of 10 BL Lac objects.
This is not yet statistically significant. The only galaxy in the sample,
OQ\,208, was not detected. More interesting, there was no correlation between
the detectability of CP and the fractional linear polarization of the core
component.

In a small survey using the VLBA at 8 and 15 GHz, we have detected CP in 3 
sources out of 11 ($\simeq 27\%$) at 8 GHz. An image of PKS 1127-114 at 8 GHz
is shown in Figure 2. The 15 GHz data are currently 
being reduced to obtain the CP spectrum.

A more sensitive ATCA survey of of 32 AGN at 5 GHz (Rayner et al. 2000) 
yielded 17 detections ($\simeq 52\%$) with (integrated) fractional
polarization levels $m_C$ from $0.03$ to $0.45\%$. 
If the 8 galaxies are excluded, only one of which was detected, the detection 
rate rises to $65\%$. (This result may be relevant to unified schemes,
e.g. Barthel 1989). Rayner et al. also
found no difference in the detection rates for quasars and BL Lac objects,
and no dependence on the fractional linear polarization of the source.
The most strongly circularly polarized source was the the intra-day 
variable (IDV) source PKS 1519-273 that varied between $-0.25$ and $-0.45\%$
on a time scale of {\em hours} (Macquart et al. 2000).

Thus it is clear that detectable circular polarization is a common phenomenon
among flat spectrum radio sources, and is associated with the most compact 
features.
It is interesting to compare recent measurements to those made
two to three decades ago (Weiler \& de Pater 1983, Komesaroff et al. 1984).
This can be done for 29 sources if we ignore mismatches in observing 
wavelength (the older measurements tended to be at longer wavelengths). We 
find that all sources that exhibited strong CP ($> 0.15\%$) in the past have
been detected recently at comparable levels, and all sources that were weak or
not detected in the past are presently also weakly polarized or not detected.
Thus despite its variability, the overall {\em level} of CP appears to be a 
persistent property of these sources. As yet we do not understand why some
sources consistently exhibit strong CP while others do not.

\section{Mechanisms for producing circular polarization}
The origin of the observed circular polarization signal is far from settled. 
One clue comes from its spectrum. It is well known that
he intrinsic circular polarization of optically thin synchrotron
radiation has a spectrum $m_C \propto \nu^{-0.5}$, where $m_C = V/I$ is the
fractional CP.
For 3C\,279 we argued in favor of Faraday conversion, based in part on 
the observed 
steep spectrum (Wardle et al. 1998). However, in other AGN
the spectrum of the integrated fractional CP is rather flat (Rayner 2000), 
while in SS433, Fender et al. (2000) found $m_C \propto \nu^{-0.9}$. It is
important to note that for a Blandford-K\"{o}nigl (1979) inhomogeneous 
jet, the CP spectrum may be quite different than for a homogeneous component
(see section 4). Coherent emission mechanisms, that may be relevant to AGN 
cores,
and especially to intra-day variable (IDV) sources can generate copious
CP, and a particularly interesting new mechanism, scintillation by a
birefringent screen, has been proposed by Macquart \& Melrose (2000).
We now briefly review these mechanisms.

\subsection{Coherent radiation mechanisms}
These have been most recently discussed  by Benford \&
Tzach (2000) and by Bingham et al. (these proceedings). Coherence may arise
from particle bunching, or from an anisotropic distribution function leading
to a negative absorption coefficient and maser action.

Both cases appear to produce rather high fractional CP (though this may be 
diluted in a real source) and are intrinsically narrow band phenomena.
The latter property can be tested. It
appears to be ruled out by the rather flat CP spectrum observed in most AGN
over widely separated wavelegths. Even for the IDV source PKS 1519-273, 
the fluctuations in both I and V were strongly correlated between 4.8 and 8.6
GHz (Macquart et al., these proceedings). Also, over a much narrower 
frequency range, we find that for sources 
observed at the VLBA, the CP signal is equally present in all 4 IF bands.

\subsection{Scintillation}
A new and intriguing possibility proposed by Macquart \& Melrose (2000) is
that of scintillation by a birefringent screen, which can generate circular
polarization from the total intensity (Stokes I). In this mechanism, Faraday
rotation in the screen causes a very small displacement of the left- and 
right-hand components of the total intensity. The two amplitude patterns 
produced by scintillation are therefore also slightly displaced from each 
other. As they sweep over the antenna there is an instantaneous, fluctuating
CP signal, though the time averaged CP must of course be zero. 

Macquart \& Melrose show that the rms fractional CP can be written as 
\begin{displaymath}
m_C \simeq \frac{\Delta x}{r_{scint}} m_I,
\end{displaymath}
where $m_I$ is the fluctuation index in total intensity, $r_{scint}$ is the
linear scale of the scintillation pattern (whether diffractive or refractive)
in the observer plane, and $\Delta x$ is the relative displacement
of the right and left circularly polarized scintillation patterns in the
observer plane.

They demonstrate that birefringent scintillation in our own Galaxy 
produces an insignificant effect
for AGN, but may well contribute to the observed CP in pulsars. For AGN
(and for Sag A*) they suggest a screen close to the source. The displacement
of the two scintillation patterns at the observer is
\begin{displaymath}
\Delta x \simeq \frac{L\lambda^3}{2\pi}\bigtriangledown_r RM,
\end{displaymath}
where $L$ is the distance from the screen to the observer, and
$\bigtriangledown_r RM$ is the transverse gradient in Rotation Measure at the
screen. For significant scintillation, they also require that $\Delta x/L
> \theta_s$, where $\theta_s$ is the angular size of the source. These lead to
rather stringent requirements on the screen, and also suggest that it should 
be mainly a long wavelength phenomenon. Macquart and Melrose apply this 
mechanism to the 
quasar 3C\,345, and find they require a RM gradient of about $10^6$ 
rad m$^{-2}$ pc$^{-1}$. So far, typical RM gradients measured in quasar cores
are of the order of a few hundred rad m$^{-2}$ pc$^{-1}$ (Taylor 1998), though
in the innermost 0.1 pc or less, it could certainly be much higher. A second 
problem is that
such a RM gradient would almost certainly completely depolarise a typical
VLBI core or inner component. This might not be a problem for an IDV source
such as PKS 1519-273, where the inferred source size is extremely small, or
for Sag A*, which exhibits no detectable linear polarization at any wavelength.

The timescales they derive for scintillation induced CP are very short: a few 
minutes for diffractive scintillation, and several hours for refractive 
scintillation. The {\em sign} of the induced CP should reverse on these
timescales. These are much too short for the observations of most AGN, but 
again are not a problem for IDVs or Sag A*. It is also possible that a 
different source-screen geometry might lead to much longer timescales.

\subsection{Intrinsic circular polarization}

Synchrotron radiation has a small intrinsic component of circular polarization
(Legg \& Westfold 1968). In a uniform magnetic field,
the maximum fractional circular polarization at frequency $\nu$ is $|m_{C}| 
\approx 1/\gamma$, where $\gamma$ is
the Lorentz factor of electrons radiating at $\nu$. More realistically, for a
combination of a uniform and a tangled field, we can write
\begin{displaymath}
m_{C} = -1.6 \, \Lambda \,(\nu/\nu_{B\perp})^{-\frac{1}{2}}
 (B_u /B_{\perp})\sin\epsilon,
\end{displaymath} 
where $B_u$ is the uniform component of the magnetic field (assumed here
to be parallel to the jet), $B_{\perp}$
is the r.m.s. value of the magnetic field in the plane of the sky,
$\nu_{B\perp} = eB_{\perp}/2\pi mc$ is the electron gyro frequency, 
and $\epsilon$ is the angle between the jet normal and the line of sight
in the frame of the emitting fluid ($\cos\epsilon = \delta \sin\theta$).
The factor $\Lambda$ accounts for the reduction in circular polarization
if $B_z$ is not also unidirectional (i.e. if there are
reversals in direction of $B_u$), and if the charges of the radiating
particles are not all the same sign. Defining 
$f_B = |\int \! B_{u}\, dl|/\int \, |B_{u}|\,dl$, and 
$f_C = (n^- - n^+)/(n^- + n^+)$, where $n^-$ and  $n^+$ are the electron and
positron densities respectively, then $\Lambda = f_B f_C$. The Faraday 
depth is also reduced by the same factor. 

In a real jet, even if the component of magnetic
field parallel to the jet is unidirectional at its base ($f_B \sim 1$), 
it will decay faster with radius ($\propto r^{-2}$) than
the transverse field
components ($\propto r^{-1}$). Further down the jet, the parallel component 
of field is likely to be dominated by sheared loops of field (e.g.
Begelman, Blandford \& Rees 1984) and $f_B$ 
(and hence $\Lambda$) is likely to be small, regardless of whether the jet is
composed of electrons and positrons or electrons and protons.

This factor has generally been neglected in the literature, and it should not 
be. It applies both to the interpretation of circular polarization, and to
the interpretation of Faraday rotation. It is sometimes assumed that
the value of $(B_u /B_{\perp})$ is indicated by the degree of linear 
polarization, but this may be quite misleading. For instance, in a Laing sheet
topolgy (Laing 1980), the fractional linear polarization may approach $70\%$,
but internal Faraday rotation and the intrinsic component of circular
polarization will both be zero.

\subsection{Faraday conversion}
A less well known but equally important mechanism for generating circular
polarization is Faraday conversion (Jones \& O'Dell 1977; Jones 1988). 
This is based
on the fact that the normal modes for radiative transfer 
in an anisotropic plasma are not purely
circular (which leads to Faraday rotation), but slightly elliptical. The
small component of linear birefringence converts Stokes parameter U to V,
and {\em vice versa}. Both rotation and conversion are caused by 
the lowest energy
relativistic electrons, and therefore serve as a probe of the low energy end
of the electron energy distribution, which is otherwise unobservable. 
An important
difference between them is that Faraday rotation is proportional to the
electron gyro frequency, $\nu_{B\perp}=eB_{\perp}/2\pi mc$, and hence to 
the sign of the 
charge on the electrons, while Faraday conversion is proportional 
to $\nu_{B\perp}^{2} \propto e^2 B^2$. Thus an
equal mixture of electrons and positrons ($f_C = 0$) can produce Faraday 
conversion,
but not rotation, as can sheared loops of magnetic field ($f_B = 0$).

It is also important to note that Faraday conversion acts on Stokes U,
while the synchrotron mechanism produces only Stokes Q (in the local frame
of the magnetic field). Stokes U can be
produced stochastically by a tangled field, or, more efficiently, by 
internal Faraday rotation.
At small optical and Faraday depths, the fractional circular polarization 
produced by conversion is
$m_C \approx  \frac{1}{6}\tau_{F}\tau_{C} m_{L}^{2}$, where $\tau_{F}$ 
and $\tau_{C}$
are the Faraday depth and ``conversion depth'' respectively, and 
$m_{L}$ is the fractional linear polarization. For a power law
distribution of electron Lorentz factors, $n(\gamma) \propto 
\gamma^{-2}$ (corresponding to $\alpha = 0.5$), with a low energy cutoff at 
$\gamma = \gamma_{min}$, $\tau_F$ and $\tau_C$ can be written as (Jones
\& O'Dell 1977)
\begin{eqnarray*}
\tau_F & \approx & 1.27 \,\tau \,\Lambda\left(\frac{\gamma}{\gamma_{min}}
\right)^{2}
\frac{\ln\gamma_{min}}{\gamma_{min}}
\frac{B_u}{B_{\perp}}\sin\epsilon \\
\tau_C & = & -0.96 \,\tau \ln (\gamma/\gamma_{min})
\end{eqnarray*}
\noindent where $\gamma$ is the Lorentz factor of the radiating electrons.
These expressions apply to the optically thin, Faraday thin case. 
The strong wavelength dependence of both $\tau_F$ and $\tau_C$ imply a very 
steep CP spectrum, $m_C \propto \nu^{-5}$. As the Faraday depth increases,
the effect saturates, and $m_C \propto \nu^{-1}$. At apreciable optical
depth, the spectrum flattens further. The reader should refer to Jones \& 
O'Dell (1977) for the complete expressions for a homgeneous source. Also,
at significant Faraday and conversion depth there can be nulls and sign 
reversals in the CP (since $Q$ has been converted to $-U$ etc.) along a 
single line of sight, or in a uniform slab (Kennett \& Melrose 1998). 
In a real source, where we average over many lines of sight,
the effect is more likely to lead to depolarization of both the circularly
and linearly polarized radiation.

Certain magnetic field topologies may also produce efficient conversion, 
without the need for internal Faraday rotation (e.g. Hodge 1982). For 
instance, in a helical
magnetic field, 'Stokes Q' from the back of the jet may appear as 'Stokes U' to
the front of the jet, depending on the pitch angle. The sign of the CP is 
determined by whether the helix is left handed or right handed.
A force-free magnetic 
field is another such configuration, though we have not yet calculated the
resulting CP in these cases. Since internal Faraday rotation is not invoked
here, the CP spectrum will be much less steep, and the problem of sign 
reversals much less severe.

Finally, the expression for $m_C$ at the beginning of this section contains two
factors of $m_L$, the fractional linear polarization. One factor is because
conversion acts on the linearly polarized radiation, and the second factor
expresses the conversion efficiency in a partially ordered field. One might
therefore expect a strong correlation between $m_C$ and $m_L$, but this is
not generally seen in either the ATCA or the VLBA results (though in 3C\,273,
CP did not appear until significant linear polarization also appeared in 
the core). A plausible reason for the lack of an obvious correlation
is that the CP originates from the core regions, which are thought to be
depolarized by an {\em external} Faraday screen (Cawthorne et al 1993, Taylor
1998, Wardle 1998) in the nucleus of the source.

\section{The circular polarization spectrum, and inhomogeneous jets}

In general a steep CP spectrum favors Faraday conversion (Jones \& O'Dell 
1977), where it can vary in the range $\nu^{-1}$ to $\nu^{-5}$ in 
an optically thin 
homogeneous component, while the intrinsic component has a $\nu^{-1/2}$
spectrum. Both are flattened by apreciable opacity.

But a flat spectrum can also
result for the ``core'' of a Blandford-K\"{o}nigl (1979) inhomogeneous jet, 
where the optical surface moves down the jet with increasing wavelength,
and this region is seen as the 'core.' In their canonical case ($B \propto
r^{-1}$, electron density $\propto r^{-2}$, and $\alpha = 0.5$), the location 
of the optical surface $r_{\tau = 1} \propto \nu_{obs}^{-1}$. The Lorentz
factor of the electrons radiating at $\nu_{obs}$ from that region is
$\gamma \sim \nu_{obs}/\nu_{B\perp}$ which is independent of $\nu_{obs}$.
Since the intrinsic component of CP is $\sim 1/\gamma$, it too is independent
of $\nu_{obs}$, and the CP spectrum is flat. 

Similar considerations show that Faraday conversion is also independent of
$\nu_{obs}$ (the Blandford-K\"{o}nigl model is isothermal, so 
$\gamma_{min}$ is also
constant). It follows that there will also be no sign reversals at longer 
wavelengths.

The point is that the simple spectral signatures in homgeneous sources almost
certainly will not be seen in real jets, where opacity and inhomgeneity 
cannot be ignored. Unfortunately, this probably requires numerical simulations
using the full equations of polarized transfer to calculate the expected CP. 

\section{Electron Positron Jets?}

If the dominant production mechanism for CP in AGN is Faraday conversion, then
this requires that the electron energy spectrum extend down to low energies in
order to get sufficient mode conversion from linear to circular polarization. 
The most robust
result for 3C\,279 (Wardle et al. 1998) was that over a wide range of source 
models, the low energy cut-off in the electron energy spectrum had to be below
$\gamma_{min} = 20$, and could be much lower. We then suggested that the 
radiating particles were 
predominantly an electron-positron pair plasma so that the jet did not 
carry far more kinetic energy than is observed to be dissipated. 

This argument has also been made by
others, e.g. Reynolds et al. (1996) for M87 and Hirotani et al. (1999, 2000) 
for 3C\,279 and 3C\,345; and see Celotti \& Fabian (1993) for a different
conclusion. The argument assumes that we can correctly estimate the 
kinetic energy
carried by a jet, which is not altogether clear (c.f. various contributions in
these proceedings). However, the argument is greatly strengthened if the 
presence of low energy particles can be inferred directly from CP measurements.

It should be noted that if $\gamma_{min}$ is small enough, then the same 
low energy particles that cause Faraday
conversion will also give excessive Faraday rotation and depolarization
unless $\Lambda = f_B f_C$ is also small. 
The degeneracy between $f_B$ and $f_C$ is not easily broken, but if $f_B$
can be estimated in some other way, then CP observations could demonstrate
the presence of electron-positron pairs
(i.e. $f_C < 1$), independent of the energy argument.

Clearly we would like to apply the same analysis to other sources in which we 
have detected CP, to see if electron-positron pairs are a general
characteristic of extragalactic jets. To do so requires multi-wavelength
observations to establish the CP spectrum, and to determine the magnetic field
strength from the self-absorption turnover. Such observations are in progress.

\section{SS433 and Galactic jets}
The discovery of circular polarization in the celebrated x-ray binary SS433
(Fender et al 2000) opens the possibility of making high resolution CP
observations of
this and other Galactic jets, where the similarities and differences compared
to AGN jets are equally striking. The ATCA observations found a strong CP 
signal ($m_C \simeq 0.6\%$ at 1.5 GHz) with a spectral index 0f $-0.9$.
The CP flux density was several mJy, easily detectable with the VLBA,
and presumably the {\em local} fractional polarization is considerably higher.
The CP mechanism is unclear (and its spectrum is complicated by both 
synchrotron opacity and free-free absorption), but both Faraday conversion
and the intrinsic mechanism are possible. The latter is a candidate because
the magnetic field is stronger than is typical in AGN -- Paragi et al. 
(1999) estimate 0.4 G at the position of the 1.5 GHz core -- and the magnetic
field structure may be simpler, with fewer field reversals. If the former
applies, then it may be possible to determine whether the composition of the
radio jets is primarily pairs or baryonic.

SS433 precesses with a period of 164 days, and the geometry is well understood
(Margon \& Anderson 1989). This should produce
a clear signature in the circular polarization in both the intrinsic and
the conversion mechanisms. For most of a period, the sign of CP should be
the same in both the jet and the counter jet. But when the jet makes an angle 
of less than $75^{\circ}$ to the line of sight, its sign should reverse. This
is a straightforward consequence of relativistic aberration ($v_{jet} =
0.26c$); for part of each period we are, in effect, looking up the jet 
instead of down it. Because of the changing geometry, we have a unique
opportunity for detailed analysis of the polarization properties of these jets.
This may shed light on the physics of extragalactic jets, for which we do not
have the luxury of observing from different angles.

\section{Long-term persistence of the sign of CP}
It was pointed out in section 2 that the presence or absence of detectable
CP appears to be a persistent property of AGN over two to three decades.
Equally interesting, for the ten strongly polarized sources for which we have 
both recent and historical data, eight show the same
{\em sign} of circular polarization at both epochs. Komesaroff et al (1984)
pointed out that the sign of the CP tended to persist over the few years of 
their observations. Now it appears that the sign persists over decades
(pointed out for a smaller sample by Homan \& Wardle, 1999). This
is highly significant because it is very much longer than the characteristic
time scale of the CP variability and of outbursts in total intensity (typically
a year or so). This 
suggests that the sign is determined by a fundamental and long-lived property
of the jets. In both the Faraday conversion and the intrinsic synchrotron
mechanisms for generating the observed CP, the sign is determined by the
net {\em unidirectional} component of the magnetic field, i.e. by the 
{\em net magnetic flux} carried by the jet. It has long been realized that 
most of the magnetic field consists of sheared and stretched loops
(Begelman, Blandford \& Rees 1984), and that
the true net flux $\int {\bf \vec{B}} dA$ must be much smaller than the 
value of $B_{rms} \times A$ ($A$ is the area of the jet cross-section) 
often calculated for synchrotron emitting jets.

CP measurements may permit proper calculation of the net magnetic flux in jets,
which, assuming flux conservation, gives the magnetic flux at the central 
engine. This is a fundamental parameter in electro-magnetic models of jet 
production (e.g. Blandford \& Payne 1982; Lovelace \& Romanova 1995; Meier, 
these proceedings; Begelman, these proceedings ), and may also relate
properties of the jet to those of the central massive black hole (Rees, 1984).

\section{Conclusions}
It was stated in the introduction that
``circular polarization (CP) measurements of extragalactic radio sources have
long enjoyed the dubious reputation of being challenging to carry out and 
hard to interpret.'' This is still true. It is hoped that this review
will convince some readers that the unique information contained in such
observations makes them, nevertheless, extremely valuable and worthwhile.

\acknowledgements
This work was supported by grants from the NSF and NASA.

\end{document}